\documentclass[12pt]{iopart}
\pdfoutput=1
\usepackage{graphicx}
\usepackage[english]{babel}
\usepackage[utf8]{inputenc}
\usepackage{booktabs}

\usepackage{color}

\expandafter\let\csname equation*\endcsname\relax
\expandafter\let\csname endequation*\endcsname\relax
\usepackage{amsmath}
\usepackage{amssymb}	
\usepackage[urlcolor=blue]{hyperref}
\usepackage{cite}

\usepackage[retainorgcmds]{IEEEtrantools} 

\begin{document}
\title[]{
  Emergence of dynamic phases
  \hbox{in the presence of different
  kinds of open boundaries} \hbox{in stochastic transport with short-range
  interactions}}

\author{Hannes Nagel and Wolfhard Janke}
\address{%
  Institut f\"ur Theoretische Physik, Universit\"at Leipzig,
  Postfach 100~920, 04009~Leipzig, Germany}

\eads{%
  \mailto{Hannes.Nagel@itp.uni-leipzig.de} and \mailto{Wolfhard.Janke@itp.uni-leipzig.de}
}

\begin{abstract}
  We discuss the effects of open boundary conditions and boundary
  induced drift on condensation phenomena in the pair-factorized
  steady states transport process, a versatile model for stochastic
  transport with tunable nearest-neighbour interactions. Varying the
  specific type of the boundary implementation as well as the presence
  of a particle drift, we observe phase diagrams that are similar but
  richer compared to those of the simpler zero-range process with open
  boundary conditions. Tuning our model towards
  zero-range-process-like properties we are able to study boundary
  induced effects in the transition regime from zero-range interactions
  to short-range interactions.
  We discuss the emerging phase structure where spatially extended
  condensates can be observed at the boundaries as well as in the bulk
  system and compare it to the situation with periodic boundaries,
  where the dynamics leads to the formation of a single condensate in
  the bulk.
\end{abstract}

\maketitle

\section{Introduction}

Stochastic mass transport processes such as the asymmetric simple
exclusion process (ASEP) or the zero-range process (ZRP) proposed by
Spitzer~\cite{Spitzer1970} are simple transport models for particle
hopping aiming to improve the understanding of basic phenomena in the
dynamics of particles in driven diffusive systems. Generally, these
particles are abstract and may represent objects from the microscopic
to the macroscopic scale when combined with appropriate dynamics.
It is this relation of abstract particles and a multitude of different
kinds of dynamics that generates manifold mappings to physical
processes and phenomena. One such phenomenon that is of particular
interest to us, is the formation of particle condensates. In fact,
dynamics leading to steady states in closed, periodic systems where
particles form condensates have been studied already for the
ZRP~\cite{Spitzer1970,Evans2000,Godreche2003,Godreche2005,Grosskinsky2003,
  Beltran2010, Beltran2012, Landim2014, Evans2005} as well as for
processes with short-range
interactions~\cite{Evans2006,Waclaw2009c,Waclaw2009b}. On
inhomogeneous structures such as a star graph or scale-free networks
even the most simple dynamics of uniform hopping can lead to
condensation at the inhomogeneities~\cite{bogacz:bb-condensation,
  Waclaw07:condensation-zrp, Waclaw2009a}. On a homogeneous structure,
condensates can emerge anywhere in the system as long as the
interaction potential falls off sufficiently
fast~\cite{Evans2006}. For a general overview of stochastic transport
processes and condensation phenomena we refer the reader to the
reviews by Sch{\"u}tz~\cite{Schuetz2000} and Evans and
Wac{\l}aw~\cite{Evans2014,Evans2015} or the book by Schadschneider et
al.~\cite{Schadschneider2011}.

While the ZRP as well as the extended models can be considered to be
driven far from equilibrium, their steady state that leads to the
condensation remains the same as in equilibrium. In fact, in the case
of systems with periodic boundaries with particle conservation, they
are constructed to have this property. This is, however, not a general
property of transport processes, as can be seen in the exclusion model
of Katz, Lebowitz and Spohn~\cite{Katz1983,Katz1984} where the
stationary distribution may or may not depend on the external field
depending on the interaction parameters of the model. It is, however,
also of interest to understand the changes to the condensation process
when this steady state is broken by replacing the periodic with open
boundaries through which particles can enter or leave the system,
thereby creating a current. In general, this external drive and
current can lead to phase separation~\cite{Popkov1999}. In fact, for
the ZRP, a specific study has been performed by Levine et
al.~\cite{Levine2005}, were among other results phase separation due
to the introduced boundary drive has been observed. In this paper we
seek to extend this approach to a stochastic transport process with
short-range interactions that feature spatially extended condensates
in its steady state. This is of interest to us because, in contrast to
the ZRP, such an extended process is able to interact with the
boundary due to its non-zero interaction range.
As a consequence we are forced to discuss different types of open
boundaries to grasp their effects on possible condensate formation and
dynamic phases. Also, instead of using a simpler transport process
with short-range interactions such as proposed by Evans et
al.~\cite{Evans2006}, we decided to employ a tunable
model~\cite{Waclaw2009c,Waclaw2009b} that can be parameterized to
resemble the condensation properties of the ZRP as well as extended
condensates such as those considered in Ref.~\cite{Evans2006}. This
allows us to compare properties of this model to those of the ZRP
discussed in~\cite{Levine2005} before going into detail with different
types of boundaries. Because the short-range interactions in that
class of transport processes are strongly related to the fact that the
steady state of a closed system factorizes over pairs of adjacent
sites, we will sometimes use the term \emph{pair-factorized steady
  states (PFSS) model},\/ although with open boundaries a steady state
does not necessarily exist.
In a previous short note~\cite{Nagel2014}, we have already briefly
discussed emerging phases and effects caused by the driven open
boundaries. We did, however, consider only one specific type of open
boundaries and were severely limited by the employed numerical
method. In a recent short communication~\cite{Nagel2015EPL}, we
sketched an improved simulation setup and discussed for this special
case the phase diagram and transition dynamics between the phases in
more detail. In particular, we pointed out that not only the details
but, in fact, the very existence of phases depends on the choice of
interaction with the boundary.
We therefore would like to complete the picture with that versatile
numerical approach and an emphasis on the point that the specific
interaction details at the boundaries have significant impact on the
system's phase diagram.

The remainder of this paper is organized as follows. In the next
section we will briefly introduce the zero-range process as well as
the tunable short-range interaction stochastic transport model and
define the considered types of open boundaries. In the third section
we describe the used numerical methods and motivate our choice for a
kinetic Monte Carlo algorithm. In the fourth section we will discuss
our results, first making a comparison with the zero-range process and
then discussing emerging phases and properties in detail with
short-range interactions turned on. Finally, we summarize our findings
in the fifth section.

\section{Stochastic transport processes with open boundaries}

\begin{figure}
  \centering
  \includegraphics{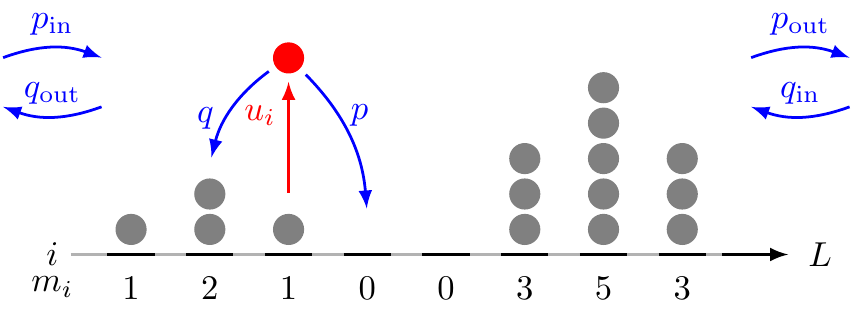}
  \caption{Schematic representation of the dynamics of a particle
    hopping process on a one-dimensional lattice with $L$ sites,
    hopping rate $u_{i}$ and drift parameters $p$, $q$. At the
    boundary sites, the drift parameters towards the boundary are
    replaced with the removal parameters $p_{\text{out}}$ and
    $q_{\text{out}}$ as indicated. Likewise, the rate of particle
    injection at the boundaries is given by the parameters
    $p_{\text{in}}$ and $q_{\text{in}}$, respectively.}
  \label{fig:zrp-scheme-open}
\end{figure}

The basic particle-hopping stochastic transport process consists of a
one-dimensional lattice with $L$ sites and a gas of $M$
indistinguishable particles. Each site $i$ of the lattice can be
occupied by any number $m_{i}=0,\ldots,M$ of particles, where
$M=\sum_{i=1}^{L}m_{i}$ is the total number of particles. These
particles can leave their sites $i$ with a rate $u_{i}$ and then jump to
one of the adjacent sites.
A target site is randomly selected among the neighbours with respect
to the strength of asymmetric hopping given by the parameters $p$ and
$q$ so that the actual rates of particles hopping to the sites right
($i+1$) and left ($i-1$) of the departure site become $pu_{i}$ and
$qu_{i}$, respectively, as indicated in
Fig.~\ref{fig:zrp-scheme-open}.
At the boundaries, particles enter or leave the system. We define
exchange rate parameters $p_{\text{in}}, q_{\textrm{in}}$ and
$p_{\textrm{out}}, q_{\textrm{out}}$ to control particle currents into
and out of the system. The exact mechanisms of injection and removal
of particles at the boundaries are discussed further below where we
define the specific properties and implementations of the boundaries.

\subsection{Zero-range process}

The hopping rate function $u_{i}$ determines the dynamics of the
particles in the system. For zero-range processes it must only depend
on the occupation number of the departure site which results in a
local-only interaction term. An example for a specific ZRP, that is of
particular interest in the context of this work, is the condensation
model with hopping rates
\begin{equation}
  u(m)=1+b/m,
  \label{eq:zrp}
\end{equation}
where a single-site particle condensate spontaneously emerges for
$b>2$ in the steady state of the periodic system when the particle
density exceeds the critical density
$\rho_{\text{crit,ZRP}}=1/(b-2)$~\cite{Bialas1997}.

\subsection{PFSS process}

In this paper, we consider a more generic model where the hopping rate
function depends also on the number of particles on the adjacent sites
as proposed by Evans et al.~\cite{Evans2006}. With an appropriate
choice of the hopping rate function, spatially extended condensates
emerge in the periodic system due to the nearest-neighbor
interaction. One generic choice of the hopping rate function of this
process reads
\begin{equation}
  u_{i} =
  \prod_{\langle i,j\rangle} u(m_{i},m_{j})
  =
  \prod_{\langle i, j\rangle}\frac{g(m_{i}-1,m_{j})}{g(m_{i},m_{j})},
  \label{eq:pfss-generic-rate}
\end{equation}
making the interaction potentials between adjacent sites
$\langle i, j\rangle$ sites isotropic for symmetric weight functions
$g(m,n)=g(n,m)$. By construction this results in a pair-factorized
steady state probability distribution of the form
\begin{equation}
  P_{M,L}(\{m\}) = Z^{-1}_{M,L}
  \prod_{\langle i, j\rangle} g(m_{i},m_{j}) \textstyle \delta_{\sum_{i=1}^{L}m_{i},M},
  \label{eq:pfss}
\end{equation}
as long as the number of particles is conserved. Here $\{m\}$ gives a
complete state, $Z_{M,L}$ normalizes the steady state similar as the
partition function in an equilibrium system and the Kronecker symbol
fixes the particle number.
  
To be able to compare to the work on ZRP with open
boundaries~\cite{Levine2005} as well as to consider a process with
effective long-range interactions, we use the hopping rate function
generated by the tunable interaction terms
\begin{equation}
  \label{eq:tunable}
  g(m,n)= \exp{ \left[
      - \left\vert m - n \right\vert^{\beta} 
      -\frac{1}{2} ( m^{\gamma} + n^{\gamma} ) \right] } 
\end{equation}
proposed by Wac{\l}aw et al.~\cite{Waclaw2009c,Waclaw2009b}. The
weights $g(m,n)$ consist of a zero-range interaction term tuned by the
parameter $\gamma$ and a nearest-neighbour interaction term tuned by
the parameter $\beta$. The steady state of this process features the
formation of particle condensates of various properties depending on
the values of these parameters. A critical density and thus
condensation phenomena exists for $0 \le \gamma \le 1$. The
condensate then assumes one out of three qualitatively distinct forms
that strongly influences the model's dynamics: a single-site peak for
$\beta < \gamma$, an extended rectangular shape for
$\gamma < \beta < 1$ or a smooth parabolic shape for
$\beta > 1$~\cite{Waclaw2009c, Waclaw2009b, Ehrenpreis2014}. Most
important to our purpose is the ability to reproduce single-site
condensates similar to those observed for the ZRP with hopping
rates~\eqref{eq:zrp} for $\beta < \gamma$ and $\gamma \le 1$, as
well as spatially extended smooth condensates for $\beta > 1$ and
$\gamma \le 1$ similar to those observed in Ref.~\cite{Evans2006}.

\subsection{Open boundaries: Mechanisms of particle exchange and external drive}

For the zero-range process the implementation of open boundaries is
straightforward because there is no interaction other than particle
exchange. For the considered model
\eqref{eq:pfss-generic-rate}--\eqref{eq:tunable}, on the other hand,
the type of interaction at the boundary sites $i=1$ and $i=L$ has to
be chosen explicitly due to its non-zero interaction range. We will
consider and discuss two main types of implementations.

Our first approach is to interpret the system as isolated and discard
the interaction terms of the bonds that cross the boundary as follows
from the factorization of weights for arbitrary graphs in
Eq.~\eqref{eq:pfss-generic-rate}. In the following, we will refer to
this type as \emph{loose} boundaries.
As a second approach we consider the system to be embedded in a larger
system with a separation that hinders particle movement like a
membrane. Here, we do not discard the interaction term for the bond
that crosses the boundary as it reflects the interaction with some
mean-field occupation outside of the considered system by setting the
external particle occupation to a constant value $m_{\infty}$. In
contrast to the first approach, we refer to this type with the term
\emph{fixed} boundaries. This results in the hopping rates
\begin{equation}
  u_{1}, u_{L} =
  \begin{cases}
    u(m_{1},m_{2}), \hphantom{u(m_{1},m_{\infty})} u(m_{L}, m_{L-1}) & \text{for \emph{loose} boundaries,} \\
    u(m_{1},m_{2})u(m_{1},m_{\infty}), u(m_{L},m_{L-1})u(m_{L},m_{\infty}) & \text{for \emph{fixed} boundaries.}
  \end{cases}
  \label{eq:hopping-loose-fixed}
\end{equation}

Additionally we consider two types of particle removal at the boundary
sites that differ in the way the hopping rate determines the rate of
particles leaving through the boundary. Intuitively, the rates of
removal are $u_{1}q_{\text{out}}$ at the first and
$u_{L}p_{\text{out}}$ at the last site. This mechanism is used for the
ZRP in Ref.~\cite{Levine2005} as well as in our own prior
study~\cite{Nagel2014}. Because normal hopping is involved for
particles leaving the system, we use the term \emph{hopping} removal
to refer to this. Here, we also consider a second removal mechanism
for particles that is symmetric to the mechanism of particles entering
the system that occurs at a constant rate. For the latter mechanism we
use the term \emph{constant} removal. Because at the boundary sites
the rate of particles leaving the system is decoupled from the hopping
rate $u_{1}$ and $u_{L}$, we replace them with $u^{\ast}_{1}$ and
$u^{\ast}_{L}$ and define
\begin{equation}
  u^{\ast}_{1}, u^{\ast}_{L} = \begin{cases}
    u_{1}, u_{L} & \text{for \emph{hopping} removal}, \\
    1,\;1 & \text{for \emph{constant} removal},
  \end{cases}
  \label{eq:removal-rates}
\end{equation}
so that the removal rates become $u^{\ast}_{1}q_{\text{out}}$ for the
first and $u^{\ast}_{L}p_{\text{out}}$ for the last site. Rates of
particles moving towards the bulk system remain unchanged ($u_{1}p$
and $u_{L}q$, respectively). 

In the following, we will only consider exchange rates that in general
reflect the external drive of the system. That is, for symmetric
dynamics ($p=q=1/2$) the exchange parameters are identical at both
boundaries
($p_{\text{in}}=q_{\text{in}}, p_{\text{out}}=q_{\text{out}}$), while
for totally asymmetric hopping ($p=1, q=0$) the exchange is restricted
to that spatial direction as well ($q_{\text{in}}=q_{\text{out}}=0$).

\section{Numerical simulation methods}

The usual approach to simulating the dynamics of a stochastic
transport process as a Markov chain is as follows: first propose a
random departure site $i$, second compute the acceptance probability
for the hop from the hopping rate $u_{i}$ and third decide whether a
particle hops to a randomly choosen neighbour. To compute an
acceptance probability it is required, however, that the hopping rate
function can be normalized for any permissible local combination of
occupation numbers. This normalization basically results in a change
of the simulation time scale by the normalization factor.

While this is possible both for hopping rate functions with an upper
bound, such as Eq.~\eqref{eq:zrp} and those proposed by Evans et
al.~\cite{Evans2006}, or when a maximum rate is known due to
conservation of the total number of particles $M(t)=\text{const}$, it
becomes inefficient for increasingly separated intrinsic time scales
of slow and fast events and thus results in a large ratio of rejected
updates.

The hopping rates of the model considered here, however, do not have
an upper bound in the regime $\beta>1$. In fact, the required
normalization constant would grow roughly as the square of the number
of particles in the system, so that the approach to directly simulate
the dynamics as sketched above cannot be used here.
To work around this limitation as well as to improve efficiency in the
presence of fast and slow events we employ a rejection free kinetic
Monte Carlo (KMC) algorithm introduced as the \emph{direct method} by
Gillespie~\cite{Gillespie1976, Gillespie1977} for the simulation of
coupled rate equations. While the method was designed for small
chemical systems with few reactions, it can be made fit for efficient
simulation of larger systems with some optimizations. The idea of the
method is similar to those of other rejection-free KMC methods such as
the $n$-fold~way algorithm~\cite{Bortz1975,Chatterjee2007} in that an
update consists of selecting an event according to its specific rate
$\Gamma_{k}$ relative to the total rate $\Gamma=\sum_{k}\Gamma_{k}$ of
all possible events, execute it and update the system time
$t\to t+\Delta t$, where $\Delta t$ is the waiting time to generate
this event. Finally the list of possible events and their assigned
rates $\Gamma_{k}$ are updated to reflect the new state of the system.
In the direct method, the event $k$ is picked by the relation
$\sum_{i=1}^{k-1}\Gamma_{i} \le \Gamma x_{1} <
\sum_{i=1}^{k}\Gamma_{i} $
and the exponentially distributed time increment $\Delta t$ is
determined as $\Delta t = -(\ln x_{2}) / \Gamma$ using two uniformly
distributed random numbers $x_{1}, x_{2}\in\left[0,1\right)$.
For our model these events are all the particle transfers between
adjacent sites $\langle i,j\rangle$ with their rates $u_{i}p, u_{i}q$
for $1<i<L$, $u_{1}p$, $u_{L}q$ and the removal and injection of
particles at the boundaries with rates
$u_{1}q_{\text{out}}, u_{L}p_{\text{out}}$ and
$p_{\text{in}}, q_{\text{in}}$, respectively.

The search for the appropriate event is easily improved by using a
binary search tree~\cite{Schulze2002} combined with multiple levels of
search, where events that originate from the same site are grouped in
the first search level, and then resolving to one specific event for
that site. The step to update the rates of events is efficiently
implemented by taking into account which events' rates actually need
updating based on which neighbouring sites were involved in the last
step of the Markov chain. This is basically an application of the
proposed update principle of the next-reaction
method~\cite{Gibson2000} which involves building dependency graphs
between events and rate recalculation to achieve this.
As an additional advantage to simpler methods, the time scale of
simulations becomes equivalent to the physical time scale, thus making
it obsolete to define artificial time scales in terms of sweeps or
local updates.

Because of the continuous time simulation method, we cannot compare
CPU time per full update, but per unit physical time of the simulated
system. For a lattice size of $L=256$ sites, the simple simulation for
the ZRP takes $31\mu s$, somewhat faster than the KMC method with
$37\mu s$ for one unit of model time. In a situation with slow and
fast events, however, the simple method becomes slower proportional to
the ratio of large to small rates, while the performance of the KMC
method does not suffer from that. In our simulations the typical
inprovement factor with only $M \approx 100$ would be around $50$ but
growing roughly with the square of the total particle number.

To compute the observables for the phase diagrams, we simulated at
least 25 replicas for each point $(p_{\text{in}},p_{\text{out}})$, and
between 50 and 100 for points near the transition lines.

\section{Results}
\subsection{Open boundary effects in the zero-range process like regime}

We start the discussion of boundary drive induced dynamical phases
with a look at the zero-range process with hopping
rates~\eqref{eq:zrp}. Most notably, for $b>2$, it features spontaneous
symmetry breaking and the formation of a single-site particle
condensate in its steady state for periodic boundaries. That is, a
single site contains a finite fraction $1-\rho_{\text{crit}}/\rho$ of
all particles, where $\rho_{\text{crit}}=1/(b-2)$ is the critical
density that is assumed on average in the rest of the system. Effects
of open, driven boundaries on this model have been studied and
discussed by Levine et al.~\cite{Levine2005}. For the ZRP, we will use
the parameter $b=5$ that results in a critical density of
$\rho_{\text{crit}}=1/3$. In the following we tune the coupling
parameters $\beta$ and $\gamma$ of our model Eq.~\eqref{eq:tunable} to
a regime where the condensation process with periodic boundary
conditions has similar properties in the steady state as the ZRP and
compare some of the properties to those found for the ZRP. With the
choice of parameters $\beta=0.4$ and $\gamma=0.6$ in
Eq.~\eqref{eq:tunable} a single-site condensate and critical density
of $\rho_{\text{crit}}=0.302 \pm 0.006$ similar to that of the ZRP is
expected for periodic boundary conditions~\cite{Ehrenpreis2014}. We
use \emph{loose} boundaries with \emph{hopping} removal, the latter of
which is the same as that used for the ZRP in~\cite{Levine2005}.

Two major phases are expected for the ZRP with open
boundaries~\cite{Levine2005}. First, a steady state with a thin
homogenous particle gas, and second, a phase with aggregate
condensates formed at one or both boundaries that act as particle
reservoirs and can influence the bulk system in between. To
distinguish between these phases we measure time series of the total
number of particles $M(t)$ and the bulk density
$\rho_{\text{bulk}}$. It is useful to determine a scaling exponent
$\alpha$ for $M(t)$ assuming it roughly follows a power-law
$M(t)\sim t^{\alpha}$ to find the two phases. The bulk density is
estimated as the average particle density in the bulk of the system
\begin{equation}
  \label{eq:bulk-density}
  \rho_{\text{bulk}} =
  \frac{1}{1 + i_{\text{bl}} - i_{\text{bf}}} \sum_{i=i_{\text{bf}}}^{i_{\text{bl}}} m_{i},
  \text{ where } m_{i}>0 \; \forall \; i < i_{\text{bf}} \;\text{and}\; i > i_{\text{bl}},
\end{equation}
that is, the region from the first ($i_{\text{bf}}$) to the last
($i_{\text{bl}}$) unoccupied site in the system.

\begin{figure}
  \includegraphics{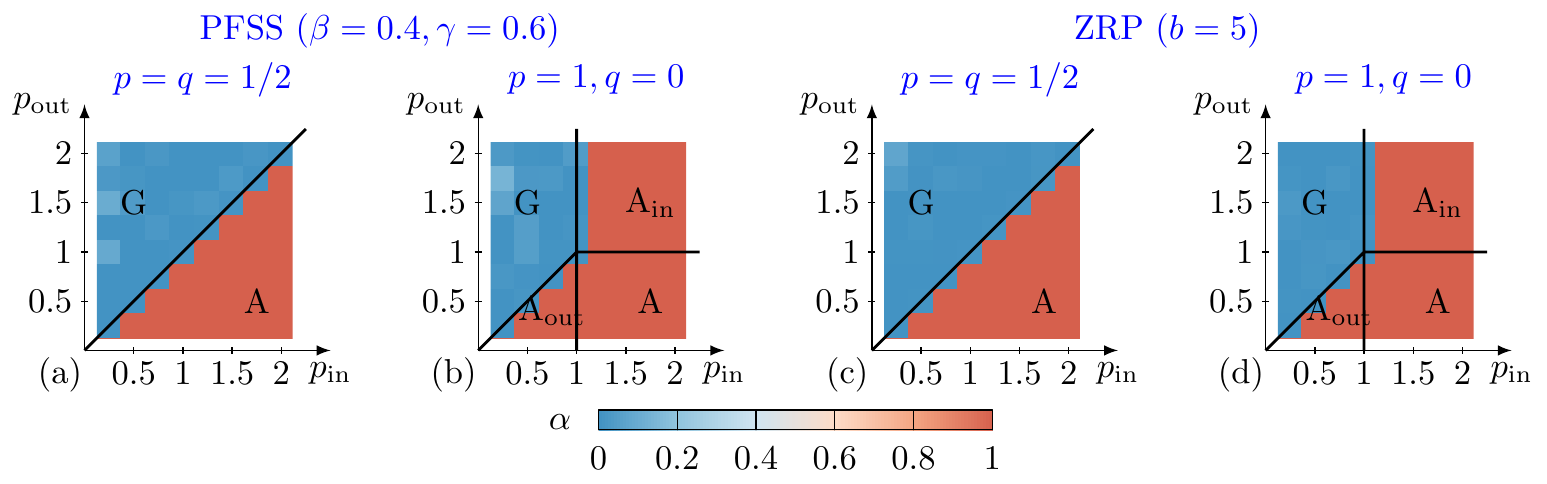}
  \caption{Scaling parameter $\alpha$ of the total number of particles
    in the ZRP-like regime for $\beta=0.4, \gamma=0.6$ and the ZRP for
    $b=5$. Low values of $\alpha \approx 0$ indicate a stable,
    fluctuating total number of particles while $\alpha=1$ shows
    linear growth in time. 
    Diagrams (a), (b) correspond to the ZRP-like model, (c),(d) to the ZRP with symmetric and totally asymmetric dynamics respectively.}

  \label{fig:zrp-totalmass}
\end{figure}

\begin{figure}
  \includegraphics{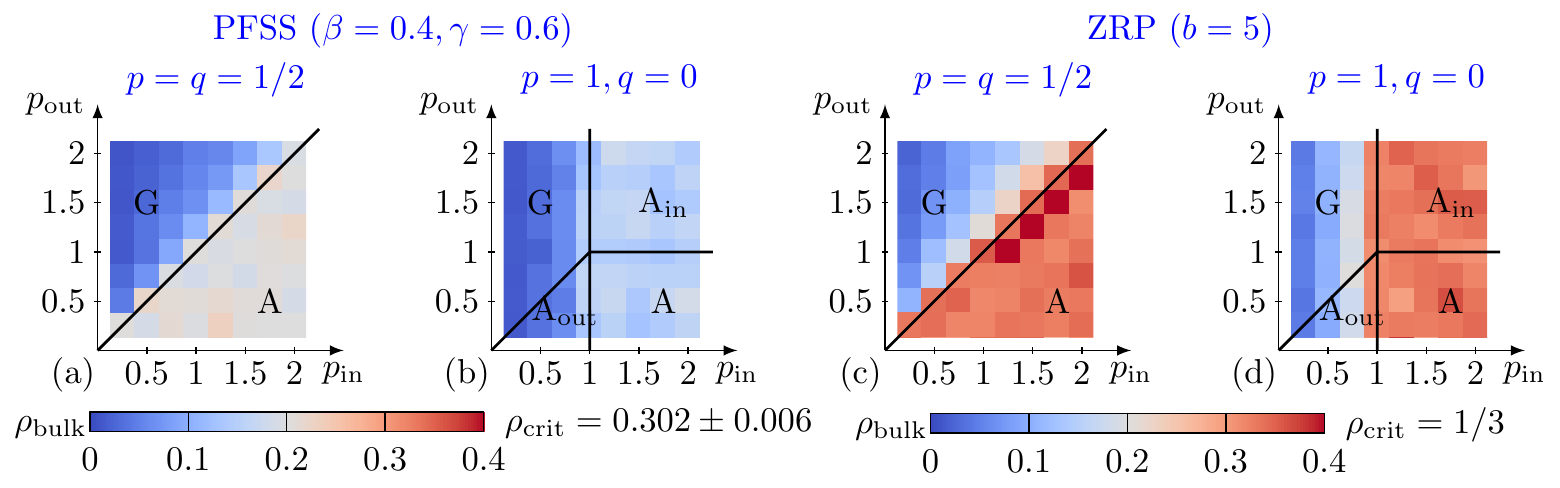}
  \caption{Bulk density $\rho_{\text{bulk}}$ in the ZRP-like regime
    for $\beta=0.4, \gamma=0.6$ and the ZRP for $b=5$. The critical
    densities for the PFSS and ZRP are
    $\rho_{\text{crit,PFSS}}=0.302 \pm 0.006$ and
    $\rho_{\text{crit,ZRP}}=1/3$, respectively. Except for the
    non-criticality of the bulk system, the dynamical phase diagram
    of the tunable system is to a high degree similar to that of the
    ZRP.}
  \label{fig:zrp-bulkdensity}
\end{figure}

From the plots of the scaling exponent $\alpha$ given in
Fig.~\ref{fig:zrp-totalmass} as well as the bulk density
$\rho_{\text{bulk}}$ given in Fig.~\ref{fig:zrp-bulkdensity} we can
clearly identify the same phases for this regime of our tunable
model~\eqref{eq:tunable} as of the ZRP. In both models there is a
particle gas phase (G) with a low stationary particle density
$\rho\equiv\rho_{\text{bulk}}$ that increases towards the transition
line to the aggregate condensate phase (A).  There, large numbers of
particles aggregate at the first and/or last sites of the
system. While the latter phase is homogeneous in systems with
symmetric hopping ($p=q=1/2$), sub-phases $\text{A}_{\text{in}}$,
$\text{A}_{\text{out}}$ and $\text{A}$, where the aggregate condensate
forms at the first site $i=1$, the last site $i=L$, or both, can be
identified for asymmetric hopping ($p \neq q$), see
Figs.~\ref{fig:zrp-totalmass} (b), (d), and \ref{fig:zrp-bulkdensity}
(b), (d). The aggregate condensates at the boundary sites act as
reservoirs for particles entering ($\text{A}_{\text{in}}$) and leaving
($\text{A}_{\text{out}}$) the system, effectively regulating particle
flux through these sites. This can be seen well for totally asymmetric
hopping, where in $\text{A}_{\text{in}}$ the bulk density assumes the
value $\rho_{\text{bulk}}=0.15\pm 0.04 < \rho_{\text{crit}}$ in the
tunable model and $\rho_{\text{bulk}} \approx \rho_{\text{crit}}=1/3$
for the ZRP. In $\text{A}_{\text{out}}$ the reservoir cannot act on
the bulk system, so that the bulk system is still a particle gas. For
symmetric hopping, however, both phases combine and aggregate
condensates at both boundaries act on the bulk system, increasing its
density to $\rho_{\text{bulk}}=0.20\pm 0.04$ for the tunable model,
still below criticality. The bulk system of the ZRP remains critical
and long-lived bulk condensates sometimes emerge, which results in
large values of the bulk density as shown in
Fig.~\ref{fig:zrp-bulkdensity}(c).

To understand the formation of a condensate at the influx boundary for
totally asymmetric hopping a simple biased random walk in the
occupation number of the first site can be
considered~\cite{Levine2005}. Because the drift
$p_{\text{in}} - (1+b/m)$ of the walker becomes positive for influx
rates $p_{\text{in}} > 1$ and sufficiently high occupation of the
first site, a stable condensate can emerge at that site. In fact, we
observe a power-law dependence of the waiting time on the influx rate
$p_{\text{in}}$ until the $\text{A}_{\text{in}}$-condensate forms
after a quench to the aggregate condensate phase. This corresponds to
the first-passage time of that random walk process to a sufficiently
high occupation number where it has positive drift. For symmetric
hopping, the argument is similar but results in a diagonal transition
line to the aggregate condensate phases for
$p_{\text{in}} \ge p_{\text{out}}$ because particles may leave
directly after entering. For a more detailed discussion of this
argument, also with respect to partially asymmetric hopping for the
ZRP, we refer to the original work of Levine et al.~\cite{Levine2005}.

The same argument can be applied to our
model~\eqref{eq:pfss-generic-rate}--\eqref{eq:tunable} in the regime
$\beta<1$, when the weak short-range interactions are negligible. The
resulting hopping rate at the first site becomes
\begin{equation}
  u_{1} = \exp\left[ 
    \frac{1}{2} (m_{1}^{\gamma} - (m_{1}-1)^{\gamma}) 
    + \vert m_{1} \vert^{\beta} - \vert m_{1} - 1\vert^{\beta}\right],
\end{equation}
approximated by
\begin{equation}
  u_{1} = \exp\left[ \frac{1}{2} \gamma m_{1}^{\gamma - 1} + \beta m_{1}^{\beta - 1} \right]
\end{equation}
for large values of $m_{1}$. This in turn approaches the value 1 for
large occupation numbers $m_{1}$ as long as $\beta < 1$, that is
for the entire single-site and rectangular condensate regimes of the
model. The drift of the first site's occupation becomes positive for
the same value of $p_{\text{in}}\ge 1$ and yields therefore the same
transition line as for the ZRP.

The formation of the aggregate condensate at the outflux boundary at
site $L$ is easily understood in the totally asymmetric case with a
similar argument as above.  For $p_{\text{in}}<1$ all particles
eventually reach the last site $L$. If the removal rate is smaller
than the rate of particles arriving at the site $p_{\text{out}} <
p_{\text{in}}$, the drift of the occupation number $m_{L}$ becomes
positive and an aggregate condensate emerges.

\subsection{Open boundary effects in the extended condensate regime}

The goal of this section is to identify the qualitative phase
structure of the described transport model depending on the strengths
of particle exchange at the boundaries with respect to the considered
types of boundary conditions. Within this section, the interaction
parameters of the tunable transport process Eq.~\eqref{eq:tunable} are
fixed at $\beta=1.2$ and $\gamma=0.6$, setting it into the regime of
smooth parabolic condensate shapes of the periodic
system~\cite{Ehrenpreis2014}. The critical density for these
parameters in a comparable system with ($L=256,M\approx L$) is
$\rho_{\text{crit}}\approx 0.3$ due to finite-size effects and
decreases to $0.125\pm 0.009$ in the limit of large systems.

\begin{figure}
  \centering
  \includegraphics{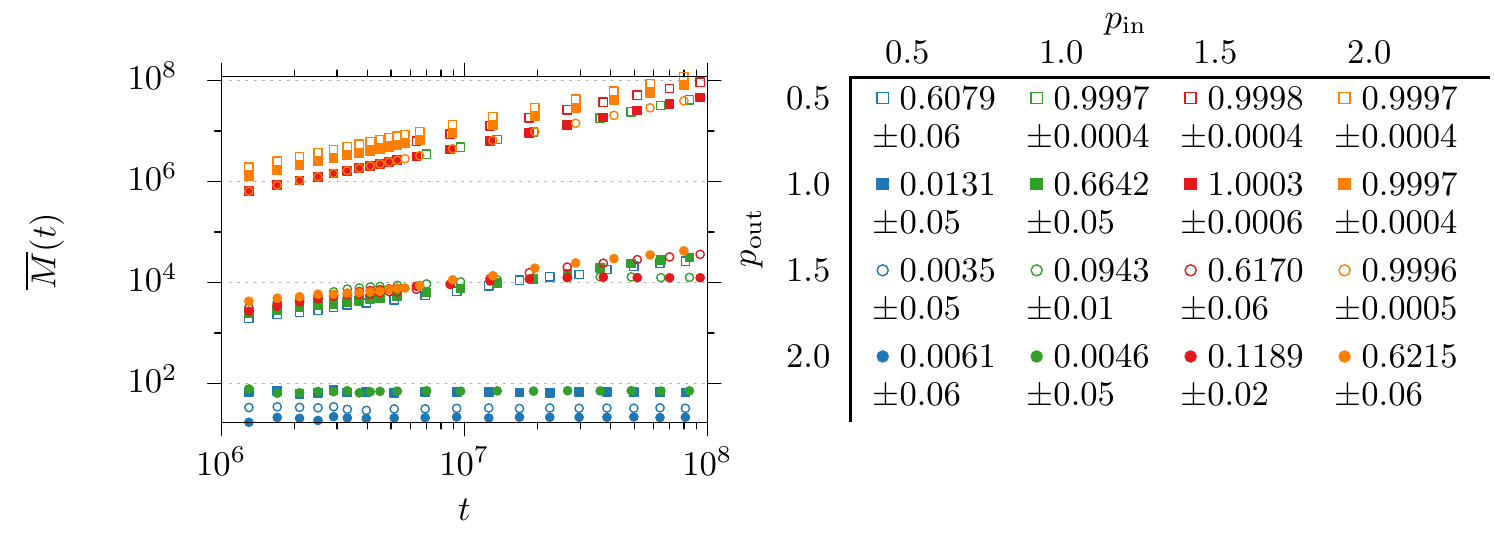}
  \caption{Average total number of particles $M(t)\sim t^{\alpha}$ for
    \emph{loose boundaries} with \emph{constant} removal of particles
    and symmetric dynamics
    ($p_{\text{in}}=q_{\text{in}}, p_{\text{out}}=q_{\text{out}}$)
    determined from 25 replicas. There are four distinct groups of
    curves: linear growth of $M(t)$ for
    $p_{\text{in}} > p_{\text{out}}$, approximate square root growth
    for $p_{\text{in}}=p_{\text{out}}$ and two groups with stationary
    particle numbers, both for $p_{\text{in}} < p_{\text{out}}$. The
    straight grey lines indicate the different observed types of
    scaling.  The scaling parameter $\alpha$ as determined from the
    average slope in the log-log-plot in the interval
    $10^{7} \le t \le 10^{8}$ is given right to the respective key
    symbol.}
    \label{fig:totalmass-plot}
\end{figure}

Based on the phase structure of the ZRP given in
Ref.~\cite{Levine2005} and numerically reproduced for the ZRP and a
short-range interaction transport model with smooth condensates in
this and our own previous work~\cite{Nagel2014} we expect to some
extent a similar phase diagram.
Therefore, to identify the phases, we continue to use the time series
of the total number of particles $M(t)$, its scaling exponent $\alpha$
and the bulk system particle density $\rho_{\text{bulk}}$ introduced
in the previous section. An example of the total mass versus time
$M(t)\sim t^{\alpha}$ for \emph{loose} boundaries and \emph{constant} removal along
with numerically determined values of $\alpha$ is given in
Fig.~\ref{fig:totalmass-plot}.
Additionally to these quantities we record the microstates of the
systems at regular intervals, so that we can compute other quantities
such as the occupation number profiles that we use later.

\begin{figure}
  \centering
  \includegraphics{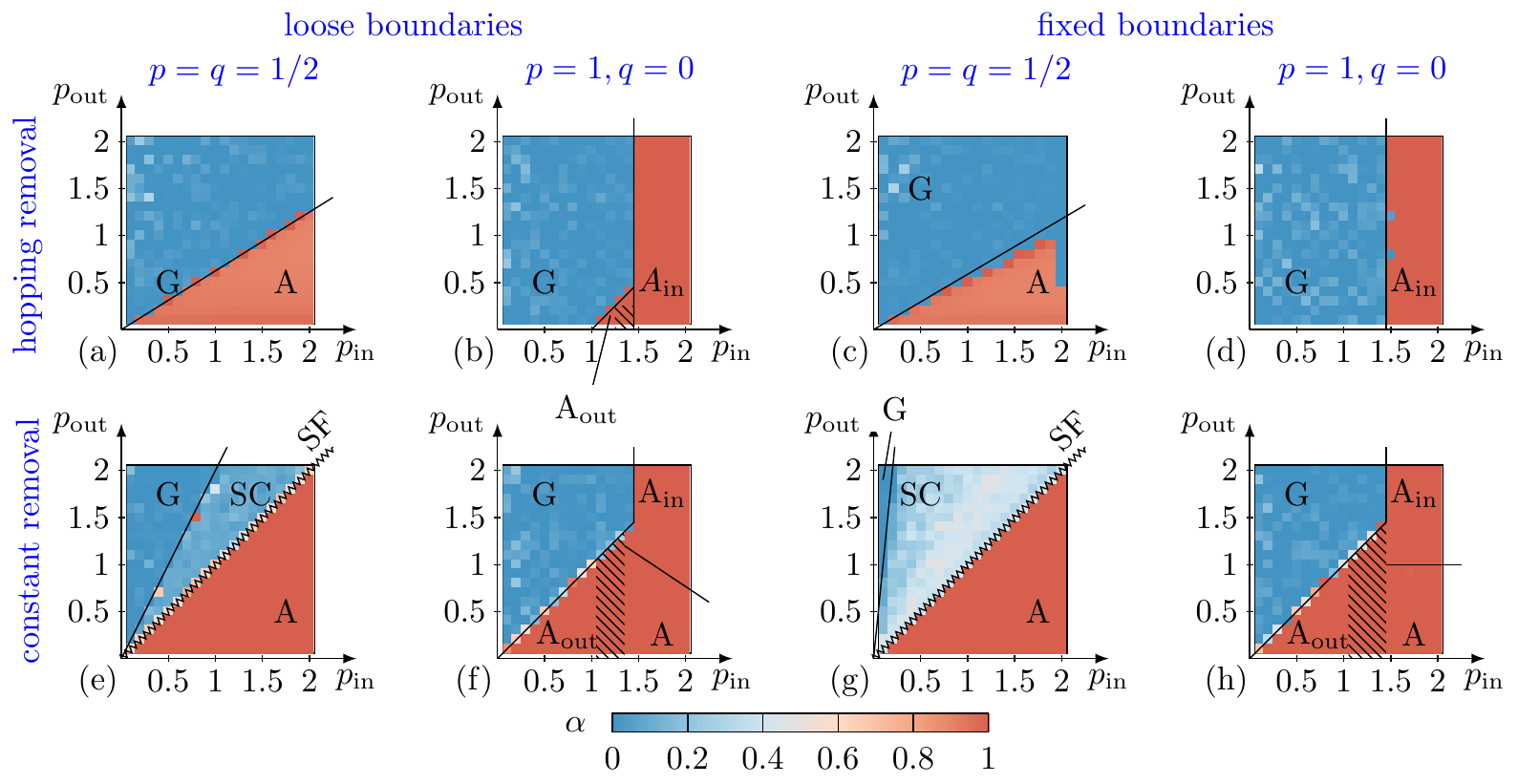}
  \caption{%
    Average scaling exponent $\alpha$ of the total number of particles
    in the system and phase boundaries for the various types of
    boundary conditions. The boundary type is given by combination of
    the labels on the left and top margins, e.g.,\ (f) \emph{loose}
    boundaries with \emph{constant} removal and totally asymmetric
    dynamics ($p=1, q=q_{\text{in}}=q_{\text{out}}=0$). The additional
    spanning condensate (SC) phase, which features a single stationary
    bulk condensate of maximal width, as well as the spanning fluid
    (SF) phase, where the system absorbs new particles, in panels (e)
    and (g) will be discussed further below in the text.}
  \label{fig:phases-totalmass}
\end{figure}

\begin{figure}
  \centering
  \includegraphics{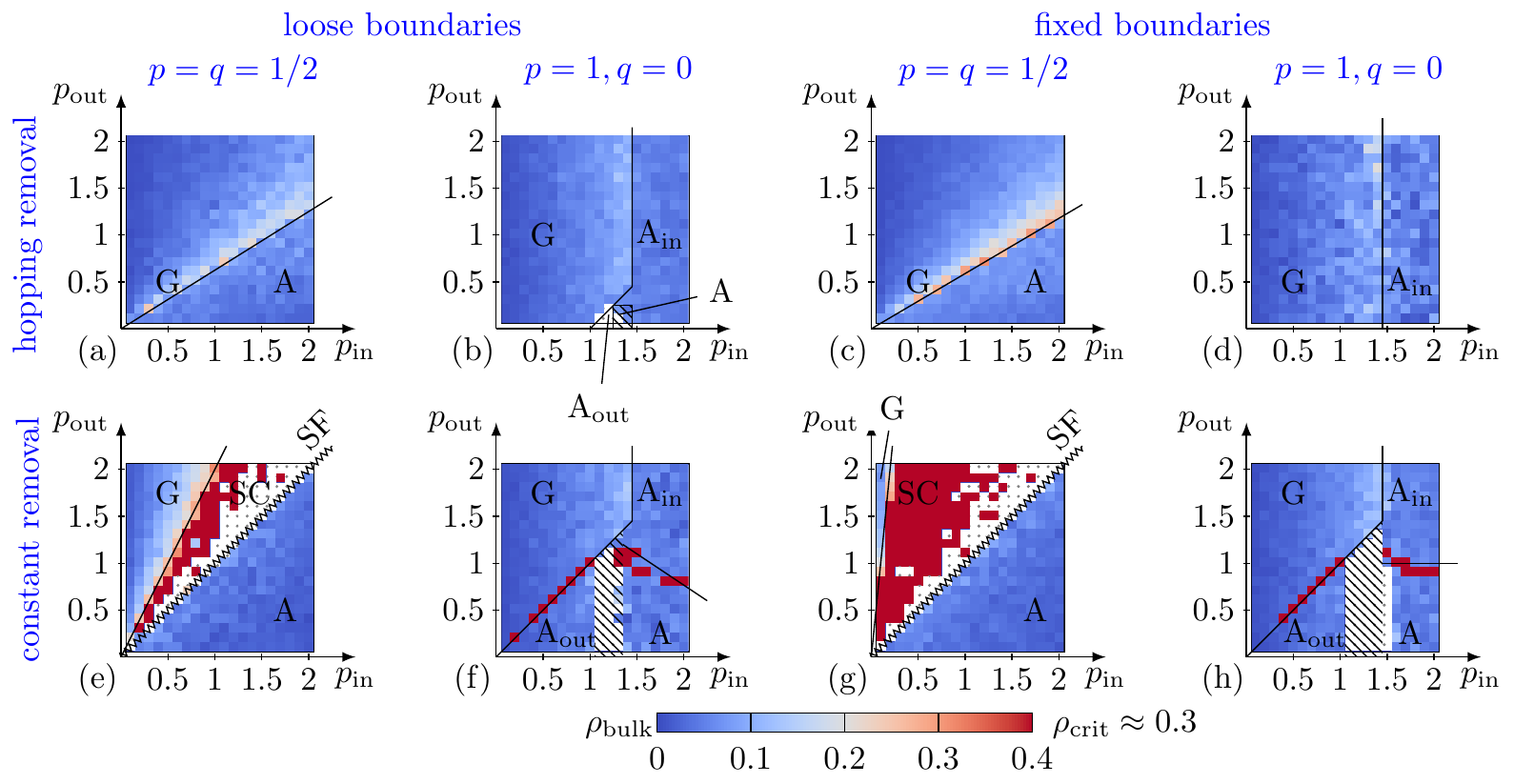}
  \caption{Particle density $\rho_{\text{bulk}}$ in the bulk system
    for given boundary drives and the four considered types of open
    boundaries. The plot values are cut off at
    $\rho_{\text{bulk}}=0.25$ to retain readability for the system
    with constant particle removal, where due to the large bulk
    condensate the density is increased by orders of magnitude. In the
    dotted regions, the bulk density remains undetermined as the bulk
    condensate has been in contact with the system boundaries in all
    simulated replicas. The hatched region in panels (b), (f) and (h)
    marks a transition region between the $\text{A}_\text{out}$ and
    $\text{A}$ phases.  The individual plots for each boundary type
    are labeled as in Fig.~\ref{fig:phases-totalmass}.}
  \label{fig:bulk-density}
\end{figure}

\begin{figure}
  (a)\hspace{-2em}
  \includegraphics[width=0.5\textwidth]{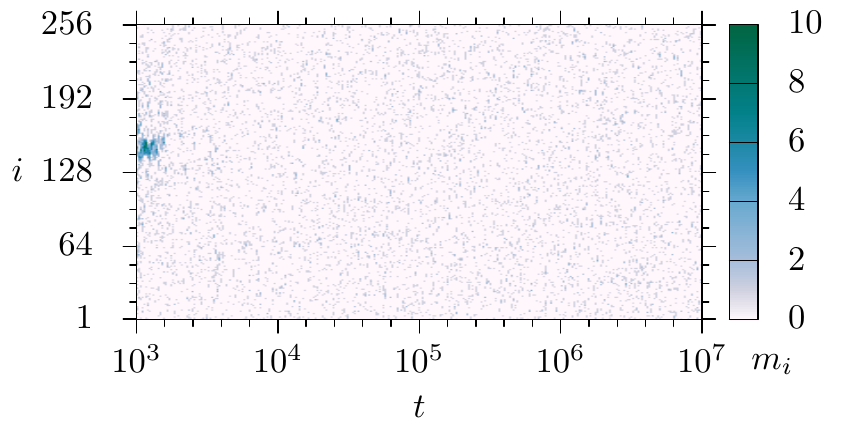}
  \hfill
  (b)\hspace{-2em}
  \includegraphics[width=0.5\textwidth]{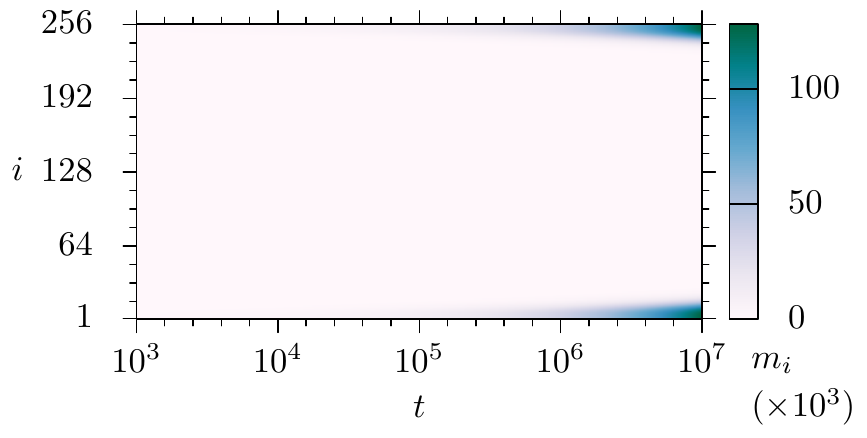}
  \\
  (c)\hspace{-2em}
  \includegraphics[width=0.5\textwidth]{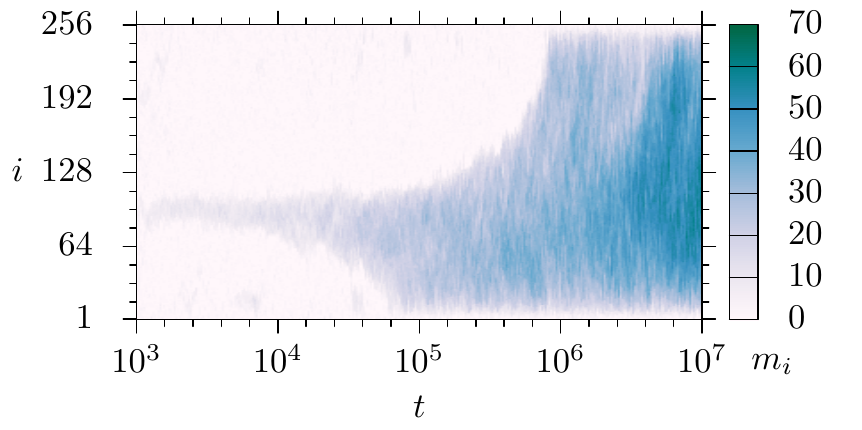}
  \hfill
  (d)\hspace{-2em}
  \includegraphics[width=0.5\textwidth]{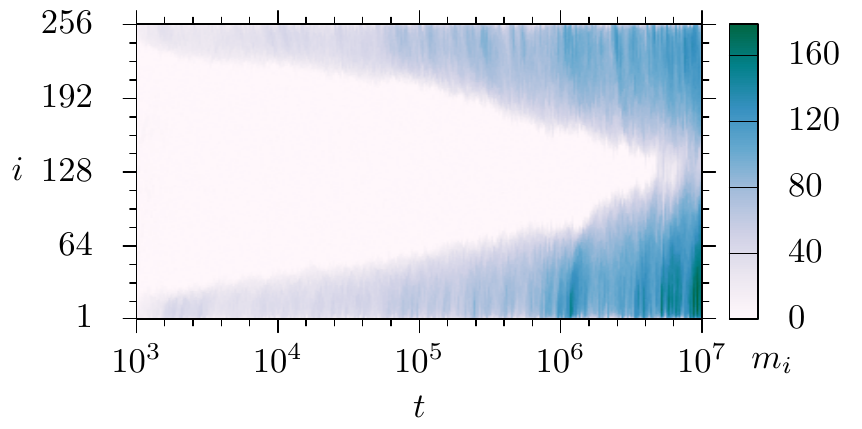}
  \caption{Example time series: (a)~Gas phase (G,
    \emph{loose/hopping}, $p=q=1/2$,
    $p_{\text{in}}=1.25, p_{\text{out}}=1$), (b)~boundary aggregate
    condensate ($\text{A}_{\text{in}}$, \emph{fixed/hopping},
    $p=1, q=0, p_{\text{in}}=p_{\text{out}}=2$), (c)~spanning bulk
    condensate (SC, \emph{loose/constant},
    $p=q, p_{\text{in}}=1.25, p_{\text{out}}=1.5$), (d) intermediate spanning fluid
    phase (SF, \emph{loose/constant},
    $p=q, p_{\text{in}}=p_{\text{out}}=2$).}
  \label{fig:snapshots}
\end{figure}

\begin{figure}
  \centering
  \includegraphics{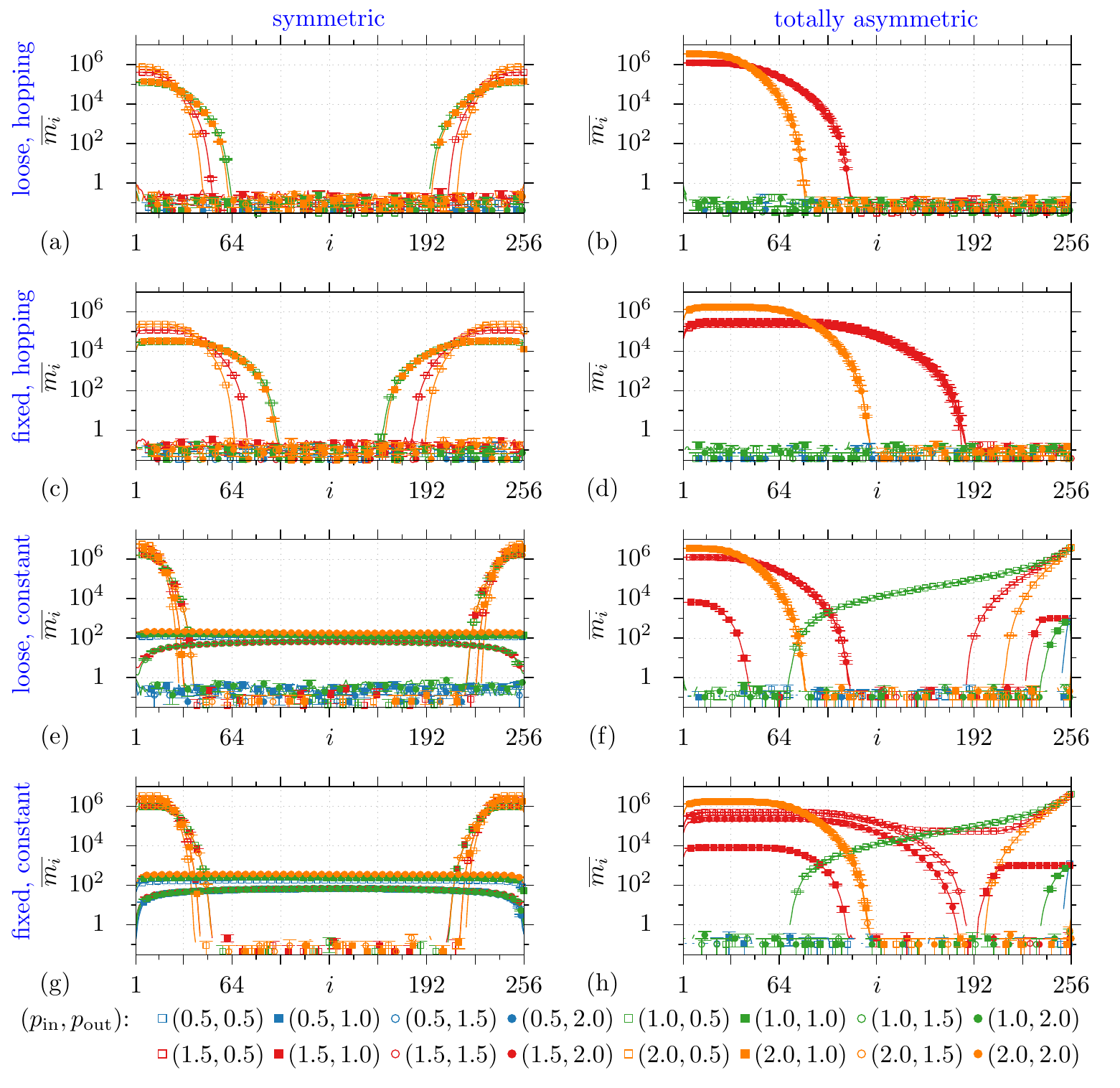}
  \caption{%
    Average occupation number profiles $m_{i}$ at simulation time
    $t=10^{8}$ for the considered types of boundary conditions. Plot
    symbols refer to influx rates, colors to outflux. The boundary
    types are from top (a,~b) \emph{loose} with \emph{hopping}
    removal, (c,~d) \emph{fixed} with \emph{hopping} removal, (e,~f)
    \emph{loose} with \emph{constant} removal and (g,~h) \emph{fixed}
    with \emph{constant} removal. The left-hand side plots represent
    results for symmetric dynamics
    ($p=q=1/2, p_{\text{in}}=q_{\text{in}},
    p_{\text{out}}=q_{\text{out}}$),
    the right-hand side results for totally asymmetric dynamics
    ($p=1,q=q_{\text{in}}=q_{\text{out}}=0$). To improve readability,
    not all points are plotted as symbols. To compute these profiles,
    25 to 40 trajectories were used.}
  \label{fig:phases-profile}
\end{figure}

As shown in Fig.~\ref{fig:phases-totalmass}, the scaling exponent
$\alpha$ identifies regions with distinct values of $\alpha \approx 0$
and $\alpha \approx 1$, that is, stationary as well as linear growing
total numbers of particles $M(t)$, respectively. For constant particle
removal, additionally the value $\alpha\approx 0.6$ is observed on the
transition line between these former regions. Together with the data
for the bulk density shown in Fig.~\ref{fig:bulk-density} we are able
to identify candidates for gas phases with low values of $\alpha=0 $
and $\rho_{\text{bulk}}$, and aggregate condensate phases where
$\alpha=1$ and low values of $\rho_{\text{bulk}}$ are
observed. Additionally a phase with stationary particle count but
relatively large bulk density is found in between those for
\emph{constant} removal and symmetric hopping. To exactly identify the
type of phase a system is in at any given parameterization
$(p_{\text{in}},p_{\text{out}})$ we use graphical representations of
the individual systems' evolution of microstates over time such as
shown in Fig.~\ref{fig:snapshots} and averaged occupation number
profiles computed from many individual trajectories shown in
Fig.~\ref{fig:phases-profile}. Combining this information we are fully
set up to identify the regions in the phase diagrams and discuss
their properties in the following subsections.

\subsubsection{Particle gas phase (G):}

\begin{figure}
  \center
  (a)\hspace{-1em}
  \includegraphics{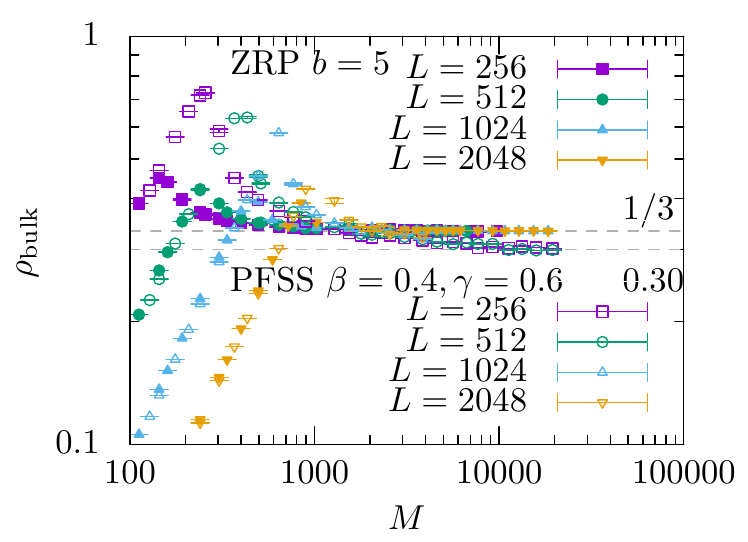}
  \hfill(b)\hspace{-1em}\includegraphics{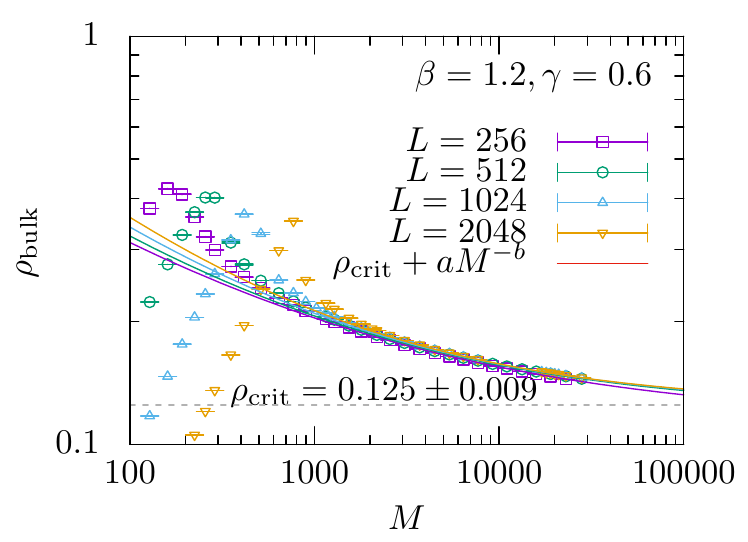}

  \caption{Finite-size effects of the critical density for low to high
    overall density in a closed periodic system of different sizes of
    $L=256, 512, 1024$ and $2048$ sites for (a) the ZRP and the
    effectively ZRP-like process with with $\beta=0.4, \gamma=0.6$ and
    (b) short-range interactions $\beta=1.2, \gamma=0.6$. For
    sufficiently large particle numbers $M$ beyond the visible local
    maxima, condensates emerge and the bulk density
    $\rho_{\text{bulk}}$ becomes the critical density
    $\rho_{\text{crit}}$. Points represent data obtained from
    simulation of the steady state ($10^{8}$ Monte Carlo sweeps),
    lines show the fitted finite-size scaling law
    $\rho_{\text{bulk}} = \rho_{\text{crit}} + aM^{-b}$ with
    $b\approx 0.40$, where the actual critical density
    $\rho_{\text{crit}}=0.125\pm0.009$ is approached for increasing
    system size $L$ and particle number $M$.}
  \label{fig:rho-critical-fss}
\end{figure}

For the considered types of interactions at the boundaries, a particle
gas phase (G) as observed for the ZRP and ZRP-like regime of the
tunable model exists. Likewise it features a thin gas of particles
filling the complete system. It is observed for small enough values of
influx rates $p_{\text{in}}$ und large enough outflux rates for
symmetric hopping $p_{\text{out}}$, so that particles can directly
enter and leave the bulk system. In the gas phase the system can be
thought of as being part of a larger periodic system. The stationary
particle density $\rho=\rho_{\text{bulk}}$ (where $\alpha=0$ as shown
in Fig.~\ref{fig:phases-totalmass}) increases with stronger drive at
the boundaries towards the critical density of the steady state system
as shown in Fig.~\ref{fig:bulk-density}. For the small system sizes we
considered to obtain most of our data, the critical density is
significantly influenced by finite-size effects. That is, for a small
total number of particles as observed in the gas phase, the critical
density $\rho_{\text{crit}}\approx 0.3$, were condensation first
occurs, is considerably larger than the large system limit
$\rho_{\text{crit}}=0.125 \pm 0.009$ which is approached with
increasing total particle number as shown in
Fig.~\ref{fig:rho-critical-fss}(a) for the ZRP and ZRP-like model and
Fig.~\ref{fig:rho-critical-fss}(b) for the extended condensate regime
of the short-range model with $\beta=1.2,\gamma=0.6$. Note that the
observed bulk density $\rho_{\text{bulk}}$ for the smaller systems is
above the asymptotic value of $\rho_{\text{crit}}$, but still below
$\rho_{\text{crit}}$ of the finite system as it should be. The values
for the critical density given in Fig.~\ref{fig:rho-critical-fss}(a)
were determined as the background density of a periodic system with
overall particle density significantly above the condensation
threshold. For a very similar ZRP with hopping rates
$u(m)=1+b/m^{\gamma}$, such finite-size effects have already been
observed~\cite{Chleboun2010,Evans2006b}.

\subsubsection{Aggregate condensate phases (A):}

For sufficiently large influx rates $p_{\text{in}}$, particles tend to
become adsorbed at the boundary and form aggregate condensates. This
is observed for any of the boundary types. As in the ZRP and the
ZRP-like regime, this phase consists actually of three regions with
aggregate condensates at the influx boundary, the outflux boundary and
at both boundaries that exist individually for totally asymmetric
hopping and mix to a single uniform region for symmetric hopping.

An example time series of an inbound aggregate condensate absorbing
entering particles is shown in Fig.~\ref{fig:snapshots}(b). The bulk
density $\rho_{\text{bulk}}$ as observed in
Fig.~\ref{fig:bulk-density} is consistently below the asymptotic value
of the critical density as determined in
Fig.~\ref{fig:rho-critical-fss}(b). The aggregate condensates show
individual shapes depending on whether they absorb inbound or outbound
particles as well as on the type of boundary (\emph{loose, fixed}) and
particle removal (\emph{hopping, constant}). The qualitative shape of
\emph{inbound} aggregate condensates $\text{A}_{\text{in}}$
[Fig.~\ref{fig:phases-profile}(b,~d,~f,~h)] closely resembles the
steady state condensate shape of the model with periodic
boundaries. The shape of the \emph{outbound} condensate
$\text{A}_{\text{out}}$ [Fig.~\ref{fig:phases-profile}(f,~h)] has a
relatively steep increase of occupation numbers towards the boundaries
but becomes almost flat when approaching the transition to $\text{A}$
with increasing influx $p_{\text{in}}$. In this transition zone
[hatched area in phase diagrams of
Fig.~\ref{fig:bulk-density}(b,~f,~h)] the aggregate condensates show
significantly increased widths with respect to mass, so that merging
of both aggregate condensates is observed very early compared to the
region $\text{A}$. As a low density bulk cannot exist, these
transition regions are also dotted in the phase diagrams of
Fig.~\ref{fig:phases-totalmass}.

The difference in condensate shapes caused by the boundary types can
be seen by comparing the respective profiles row by row in
Fig.~\ref{fig:phases-profile} and in fact is visible also in the large
bulk condensate phase where it approaches the boundary as discussed
below. For \emph{loose} boundaries, the profile starts and ends at
high occupation numbers with zero slope at the boundary. Since there
is no interaction beyond the boundary, its shape is very close to one
half of a steady state shape. With \emph{fixed} boundaries, the
condensate shape is forced to lower occupation numbers towards the
boundaries by the interaction term with the mean-field occupation
$m_{\infty}=0$. Also, the maximum occupation of the condensates
becomes lower which leads to increased condensate widths because the
total mass of the condensates is comparably independent of
\emph{loose} or \emph{fixed} boundaries.
The mechanism of particle removal at the boundaries seems to affect
the aggregate condensates only to the extent that their rates of
growth are changed. \emph{Constant} removal lets more particles escape
the system and therefore results in much smaller aggregate
condensates.

\begin{figure}
  \centering
  (a)\hspace{-1em}
  \includegraphics[width=0.5\textwidth,type=pdf,read=.pdf,ext=.pdf]{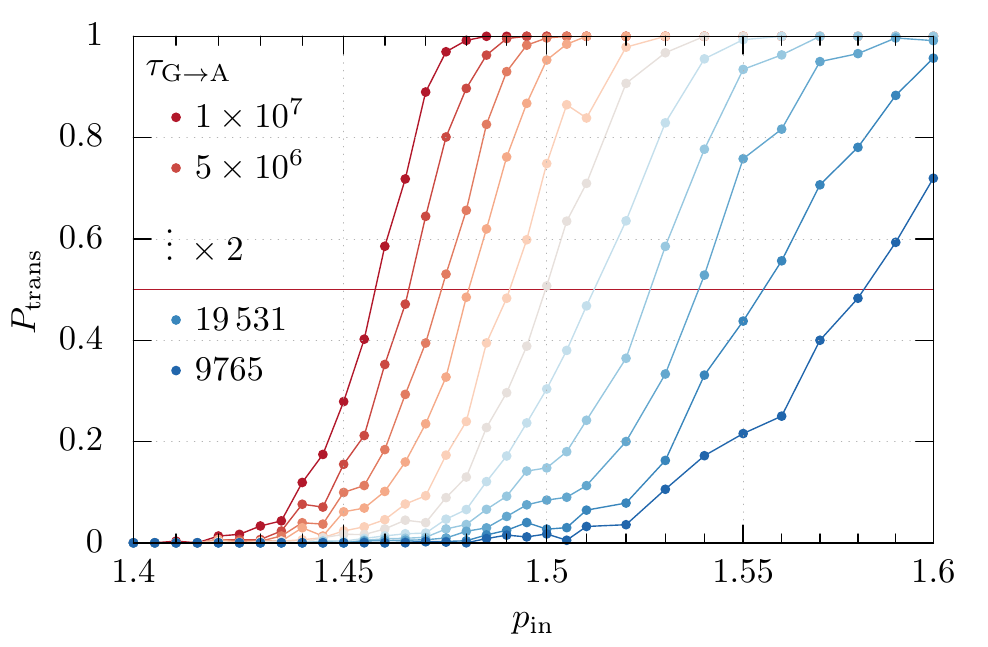}
  (b)\hspace{-1em}
  \includegraphics[width=0.43\textwidth]{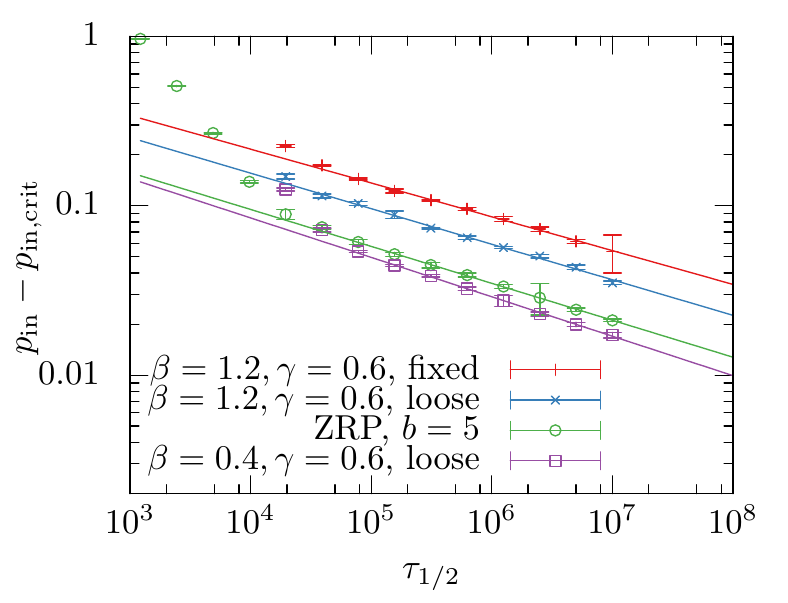}
  \caption{(a) Ratio of replicas with an emerged aggregate condensate
    after a waiting time $\tau_{\text{G} \to \text{A}}$ after a quench
    to the aggregate condensate phase~\eqref{eq:transition-ratio} for
    influx rate values of $1.4<p_{\text{in}}<1.6$ and constant outflux
    rate $p_{\text{out}}=1.5$ for totally asymmetric hopping. For this
    plot, from $N=200$ up to $800$ replicas per influx value were used
    with system size $L=256$. Larger systems give qualitatively the
    same result.  (b) Log-log plot of the excess influx rate
    $p_{\text{in}}-p_{\text{in,crit}}$ versus the half-value waiting
    time $\tau_{1/2}$, where half of the replicas have developed an
    aggregate condensate. Symbols represent numerical data determined
    from simulations of $N=200$ system replicas of size $L=256$, lines
    show the fitted scaling law~\eqref{eq:scaling-transition}.}
  \label{fig:transition-survival}
\end{figure}

To estimate the slope and position of the transition line to the phase
A, we perform quenches of a large number of replicas of systems to
several values of the influx rate $p_{\text{in}} > p_{\text{in,crit}}$
beyond the transition line, where $p_{\text{in,crit}}$ is the critical
influx rate for the given parameterization. For any value of the
``depth'' $p_{\text{in}} - p_{\text{in,crit}}$ of the quench into the
phase we then measure at several times the transition ratio of the gas
phase
\begin{equation}
  \label{eq:transition-ratio}
  P_{\text{trans}}(\tau_{\text{G}\to\text{A}}) = 
  \frac{1}{N}\sum_{i=1}^{N} H\left( M(t) - M_{\text{thresh}} \right),
\end{equation}
where $H(x)$ is the Heaviside function, $N$ is the number of replicas
for a given value of $p_{\text{in}}$ and
$M_{\text{thresh}} = a \rho_{\text{crit}} L$ is a threshold mass (with
$a>1$) to detect the transition to the aggregate condensate
phase. This transition ratio is related to the survival probability of
the gas phase as $P_{\text{surv}}(\tau)=1-P_{\text{trans}}(\tau)$. The
determined values for \emph{loose} boundaries and \emph{hopping}
  removal are given in Fig.~\ref{fig:transition-survival}(a). The
power-law scaling of the transition time becomes evident when looking
at a fixed transition ratio $P_{\text{trans}}\approx 0.5$ in
Fig.~\ref{fig:transition-survival}(a): approximately equidistant
increases of the depth of quench halve the mean waiting time to reach
that transition ratio. This is illustrated in
Fig.~\ref{fig:transition-survival}(b), which shows the scaling of the
depth of quench versus the waiting time for transition of $1/2$ of the
replicas. The values determined for this scaling relation
\begin{equation}
  \label{eq:scaling-transition}
  p_{\text{in}} - p_{\text{in,crit}} \propto \tau_{1/2}^{-\kappa}
\end{equation}
near the transition line as well as the critical influx rates
$p_{\text{in,crit}}$ are given in
Table~\ref{tab:results-transition}. Additionally to the considered
tunable model we also considered the ZRP to check our methods. 
As for the PFSS, we considered the waiting time until a condensate
attached to the influx boundary neglecting formation of droplets in
the bulk system. We only rarely observed the case were a droplet forms
away from the boundary and grew to become the aggregate condensate. We
did not observe the formation of stable condensates in the bulk. We
would like to suggest that the situation for bulk condensate formation
here is indeed different to that of the ZRP with periodic boundaries
where coarsening sets in immediately
(c.f. Ref.~\cite{Godreche2005}). The phase transition could appear,
when a droplet in the bulk grows fast enough to become immobile and
attach to the boundary instead of diffusing or leaving the
system. From our observations, the aggregate condensates in fact
formed at the boundary.

Using these observations of the scaling of transition times
  to the aggregate condensate phase, we reproduced the value of the
critical influx rate $p_{\text{in,crit}}=1$ for the ZRP (see
Ref.~\cite{Levine2005}) as well as determined the scaling exponent
$\kappa$, c.f.\ Table~\ref{tab:results-transition}.

We would also like to point out that the exponents observed for the
scaling relation of the ``depth'' of quench with the transition time
are within their statistical errors identical
($\kappa\approx0.22 \pm 0.02$), although different models (tunable
PFSS and ZRP), couplings (\emph{fixed} and \emph{loose}) or
interaction at the boundary (\emph{hopping} and \emph{constant}
  removal) are considered. The physical meaning of this scaling
exponent for a quench from the gas to the aggregate condensate phase
is the connection of the ``depth'' of the quench into the new phase to
the time it takes until it manifests in the system (or the survival
time of the old phase). The value of the scaling exponents here hints
at a universality of this transition. Possibly related to this is the
global persistence scaling exponent $\theta$, which describes the
distribution of survival times $P(\tau)\sim \tau^{-\theta}$ of the old
phase after a quench to a critical
point~\cite{Majumdar1996,Majumdar1999,Bray2013a} in non-equilibrium
systems. However, we would like to postpone this interesting question
to future work as we could not yet address it properly.

\begin{table}
  \caption{Numerically determined values for the scaling relation~\eqref{eq:scaling-transition} of the
    quench depth $p_{\text{in}}-p_{\text{in,crit}}$ versus the waiting time $\tau_{1/2}$
    to transition for $1/2$ of replicas. These values are determined for totally
    asymmetric hopping and fixed outflux rate $p_{\text{out}}=1.5$, where
    the critical influx rate does not depend on $p_{\text{out}}$.}
  \centering
  \begin{tabular}{lcc}
    \toprule
    & scaling exponent $\kappa$ & critical influx rate $p_{\text{in,crit}}$ \\
    \midrule
    $\beta=1.2,\gamma=0.6$, \emph{fixed} & $0.210 \pm 0.064$ & $1.426 \pm 0.018$ \\
    $\beta=1.2,\gamma=0.6$, \emph{loose} & $0.200 \pm 0.027$ & $1.459 \pm 0.013$ \\
    $\beta=0.4,\gamma=0.6$, \emph{loose} & $0.233 \pm 0.031$ & $1.0337 \pm 0.0038$ \\
    ZRP, $b=5$                           & $0.218 \pm 0.014$ & $1.0055 \pm 0.0024$ \\
    \bottomrule
  \end{tabular}
  \label{tab:results-transition}
\end{table}

The nature of the aggregate condensate phase is different to the
regime with negligible short-range interactions (e.g.,
$\beta=0.4, \gamma=0.6$), however. This becomes clear when looking at
the argument of the finite biased random walk of the occupation of the
boundary sites. The hopping rate at the first site with an unoccupied
neighbor does not decrease or even approach a stationary value when
the occupation number $m_{1}$ increases so that a single site
aggregate condensate cannot form. With a neighbor of similar
occupation, however, the hopping rate $u_{1}$ has a local minimum for
a non-zero occupation number, so that for this configuration of the
first sites there is a positive drift of the occupation number in the
random walk argument. Due to this interaction with the sites in the
system, a spatially extended condensate aggregates at the boundary.

\subsubsection{Spanning condensate phase (SC):}

With symmetric hopping and constant particle removal, an
additional phase featuring a single large condensate emerges
intermediate between the gas and aggregate condensate phases. The
condensate spans the bulk of the system almost approaching the
boundaries. As in the gas phase, the total number of particles $M(t)$
and therefore the condensate mass is stationary. This is already visible
from the high value for the stationary total number of particles given
in Fig.~\ref{fig:totalmass-plot} for rates $(p_{\text{in}},p_{\text{out}}) = (1.0,
1.5)$ and $(1.5,2.0)$. An example time series leading to such a
spanning bulk condensate is shown in Fig.~\ref{fig:snapshots}(c). The
resulting average occupation profile for large times is shown in
Fig.~\ref{fig:phases-profile}(e,~g). There it is visible that towards
the phase boundary to the aggregate condensate phase the bulk
condensate starts to touch the boundary sites. This becomes clear in
Fig.~\ref{fig:bulk-density}(e,~g) where the dotted region
indicates that the measurement of the bulk density as defined in
Eq.~\eqref{eq:bulk-density} cannot be achieved because the bulk does
not exist.

To understand why this additional phase occurs with the
\emph{constant} removal mechanism, we suggest that for single
particles at the boundary as it ocurs in the gas phase, the removal
rate is much lower than with the \emph{hopping} removal
mechanism. With asymmetric hopping this leads to emergence of an
outward boundary aggregate condensate ($\text{A}_{\text{out}}$) as
evident from Fig.~\ref{fig:bulk-density}(f,~h). For symmetric hopping,
however, this leads to an increase of the bulk density above the
critical density and thus the formation of a bulk condensate. This
condensate then absorbs particles until its stable maximum size is
reached. When the particle influx rate $p_{\text{in}}$ becomes large
enough to create aggregate condensates, the bulk condensate connects
to the boundaries resulting in a flat occupation profile. We mark this
transition with a zig-zag line denoted as spanning fluid (SF) phase in
Figs.~\ref{fig:phases-totalmass} and~\ref{fig:bulk-density}. The total
number of particles in this transitionary state grows roughly with
$\alpha\approx 0.6$ as shown in Figs.~\ref{fig:totalmass-plot} and
\ref{fig:phases-totalmass}.

\section{Conclusion}

We systematically studied the effects of open boundaries and external
drive in a stochastic transport process with tunable short-range
interactions~\cite{Waclaw2009c,Waclaw2009b} far from equilibrium. To
do so in a meaningful and systematic way we proposed four different
boundary types distinguished by the type of interaction and the
mechanism of particle removal at the boundary. The interaction at
\emph{loose} or \emph{fixed} boundaries reflects the non-existence or
existence of an interaction term with a mean-field occupation across
the boundary site. For leaving particles, additional to \emph{hopping}
removal, we propose \emph{constant} removal for reasons of the
symmetry of particle exchange. We considered the four types of open
boundaries generated by combinations of these properties and
determined the respective phase diagrams for both symmetric and
totally asymmetric hopping dynamics.

We successfully applied the direct KMC technique, a continous-time
Monte Carlo method to the system to control the unbounded values of
the hopping rate of the chosen dynamics for $\beta > 1$ where in the
steady state of the closed (periodic) system extended condensates of
smooth parabolic shape form. This, however, also mitigates the need to
artificially formulate update sweeps that coordinate regular particle
hops with particle injections and removals because all events can be
treated equally only according to their rates leading to a uniform
time scale.

For negligible strength of the short-range interaction
($\beta < \gamma$) we found a phase diagram that is essentially
equivalent to that of the ZRP condensation model~\eqref{eq:zrp} as
discussed in Ref.~\cite{Levine2005}. A homogeneous particle gas phase
and an aggregate condensate phase make up the phase diagram. The
transition mechanism between these can be understood the same way as
for the ZRP.

When the short-range interactions become important ($\beta>1$) we find
an enriched phase structure. The particle gas phase is identical to
that in the prior models. In the aggregate condensate phases, however,
spatially extended condensates emerge at the boundary sites with
envelope shapes that adapt to the predominant flux of particles in or
out of the system in case of asymmetric dynamics. The interaction at
\emph{loose} or \emph{fixed} boundaries, while not changing the phase
structure qualitatively, does have a significant effect on the
transition lines between phases as well as the properties in the
aggregate condensate phases. In the case of \emph{fixed} boundaries,
this is very obvious in the deformation of the aggregate condensates
at the boundary sites. With the constant rate particle removal
mechanism, however, we observed the emergence of a new intermediate
phase featuring a dominant bulk condensate between the particle gas
and aggregate condensate phases. To obtain a precise value for the
critical influx rate that separates phases G and
$\text{A}_{\text{in}}$ for totally asymmetric hopping, we analyzed
survival times of the gas state for different quenches to
$\text{A}_{\text{in}}$. An interesting observation in this analysis
was the identity of the scaling exponents of the relation between
distance to the transition line and half-value survival time across
different models, coupling strengths and considered boundary
types. While we could not yet identify the cause of this scaling, it
appears to be a universal property for this type of phase transition.
As a future project it would be interesting to work out its possible
relation to the global persistence scaling.

\section*{Acknowledgements}
We would like to thank Johannes Zierenberg for useful discussions
concerning observations at the transition to the aggregate condensate
phase and possible universal scaling there and Hildegard
Meyer-Ortmanns for fruitful cooperation in the initial phases of this
work. This project was financially supported by the DFG (German
Science Foundation) under Grant No.\ JA~483/27-1.  We further
acknowledge support by the Deutsch-Franz{\"o}sische Hochschule
(DFH-UFA) through the German-French graduate school under Grant No.\
CDFA-02-07.

\section*{References}

\providecommand{\newblock}{}

\end{document}